# Quantum Numbers of Hall Effect Skyrmions

CHETAN NAYAK[*]

*Department of Physics*
*Joseph Henry Laboratories*
*Princeton University*
*Princeton, N.J. 08544*

FRANK WILCZEK[†]

*School of Natural Sciences*
*Institute for Advanced Study*
*Olden Lane*
*Princeton, N.J. 08540*

---

[*] Research supported in part by a Fannie and John Hertz Foundation fellowship. nayak@puhep1.princeton.edu
[†] Research supported in part by DOE grant DE-FG02-90ER40542. WILCZEK@IASSNS.BITNET

# ABSTRACT

By carefully considering a family of wave functions for Skyrmions in simple quantum Hall states, whose members are labelled by a non-negative integer and which properly generalizes the traditional Laughlin quasiparticle, we argue that the spin of this particle has a fractional part related in a universal fashion to the properties of the bulk state, and propose a direct experimental test of this claim. We argue directly also for the fractional charge and fractional quantum statistics of these particles. We show that certain spin-singlet quantum Hall states can be understood as arising from primary polarized states by Skyrmion condensation.

Almost 40 years ago Skyrme [1] introduced a model of nucleons as distributions of pion fields which has inspired much work, both in its original context and more generally in the quantum theory of solitons. More than 10 years ago Wilczek and Zee [2] discussed the novel fractional spin and quantum statistics and that can arise for what they called "baby Skyrmions" in 2+1 dimensions. These objects (which we shall here call simply skyrmions) arise in an $SO(3)$ nonlinear $\sigma$-model, where they are described by field distributions of the type

$$\vec{n}(r,\phi) \; = \; (\sin\theta(r)\cos\phi, \; \sin\theta(r)\sin\phi, \; \cos\theta(r)) \;, \qquad (1)$$

where $\vec{n}$ is a unit vector field, and $\theta(r)$ runs from $-\pi$ at $r = 0$ to $0$ at $r \to \infty$. As these authors pointed out, such skyrmions arise naturally in models of ferromagnets, with $\vec{n}$ interpreted as the local direction of magnetization.

Recently there has been a revival of interest in objects of this kind, inspired by the important realization that for some quantum Hall states – including the classic $\nu = 1$ and $\nu = 1/3$ cases – the lowest energy charged quasiparticles may be skyrmions [3,4,5,6]. There is significant numerical and experimental support for these ideas.

The recent literature on skyrmions in the quantum Hall complex takes as its starting point an effective theory of the state in question which was initially postulated [4] and has since received some microscopic justification [5]. Here, by addressing the determination of quantum numbers in a more direct fashion, we refine and partially justify the effective theory. We find that the traditional Laughlin quasihole finds a natural place as a "spin zero" skyrmion. Most important, we find from our microscopic considerations that a parameter in the effective quantum theory, the coefficient of the Hopf term, is quantized, with a value displaced from integer by a universal constant depending on the bulk state. This fact is reflected in quantization in the properties of the skyrmion as material parameters are varied, and specifically to a non-integral part of its spin which is in principle observable experimentally. (Note that spin in the direction of the magnetic field, here taken as the $z$ direction, is a good quantum number. In what follows, when we refer to skyrmion spin we mean this component.) We also find that the anomalous quantum properties of the skyrmions – their fractional charge, statistics, and spin – all come together in a hierarchical construction, by way of skyrmion condensation, of quantum Hall states involving spin degrees of freedom.

## 1. Wave Function and Charge

The magnetization vector $(\sin\theta\cos\phi,\ \sin\theta\sin\phi,\ \cos\theta)$ is generated by the spinor $(\cos\frac{\theta}{2}e^{i\phi/2},\sin\frac{\theta}{2}e^{-i\phi/2})$. Thus one can generate a wave function appropriate to describe the magnetization field (1) by multiplying a standard Laughlin droplet wave function by the spinor factor

$$\prod_k \begin{pmatrix} \cos\frac{\theta(r_k)}{2}e^{i\phi_k/2} \\ \sin\frac{\theta(r_k)}{2}e^{-i\phi_k/2} \end{pmatrix},$$

where of course $(r_k, \phi_k)$ is the position of the $k$th electron in polar coordinates.

In global considerations it is desirable to avoid the complication of a boundary, so we find it convenient at this point to put the problem on a sphere. This device has been used with great success several times in the theory of the quantum Hall effect, starting with the foundational work of Haldane. The coordinates on the sphere are specified by complex numbers $\alpha, \beta$ with $|\alpha|^2 + |\beta|^2 = 1$ according to $\alpha = \cos(\theta/2)e^{\frac{i\phi}{2}}$, $\beta = \sin(\theta/2)e^{-\frac{i\phi}{2}}$, where $\theta$ and $\phi$ are the spherical coordinates. Here $0 \leq \theta \leq \pi$ and $0 \leq \phi \leq 4\pi$, and we are actually dealing with a double cover of the sphere. The resemblance between this parameterization of real space (*i.e.*, the surface of a unit sphere) and spinor space is of course not accidental, and it will play an important role in the skyrmion story.

Let us now briefly recall the rules for constructing appropriate wave functions on the sphere. They must be homogeneous polynomials in each of the coordinates $\alpha_k$, $\beta_k$ of the electrons. The degree $d$ of the polynomial reflects the common magnetic flux to which all the particles are subject, in Dirac units, according to the formula $4\pi\Phi = d$. Thus for example the Laughlin ground state at $\nu = 1/3$ is given by the wave function

$$\Psi_3 = \prod_{k<l}(\alpha_k\beta_l - \alpha_l\beta_k)^3 \ . \tag{1.1}$$

The degree of the polynomial in any coordinate is $3(N-1)$, where $N$ is the number of particles, so that the filling fraction is $N/(4\pi\Phi) = N/3(N-1) \xrightarrow[N\to\infty]{} 1/3$. A Laughlin quasihole at $(\alpha_0, \beta_0)$ is generated by multiplying $\Psi_3$ with a factor

$$f_0 = \prod_k (\alpha_k\beta_0 - \alpha_0\beta_k) \ . \tag{1.2}$$

This factor evidently raises the flux by unity without changing the number of electrons. Given that the wave-function heals in a few magnetic lengths, the modification introduced by this factor be interpreted as producing a deficit of charge $e/3$ localized around $(\alpha_0, \beta_0)$ relative to the uniform ground state.

Now for the anti-skyrmion we lift the trial wave function previously suggested for the droplet to the sphere according to

$$\Psi_{\text{skyr.}} = \Psi_3 f_{\text{skyr.}} \tag{1.3}$$

where

$$f_{\text{skyr.}} = \prod_k \binom{\alpha_k}{\beta_k} \ . \tag{1.4}$$

A wave function of this type, in the context of $\nu = 1$, appears in [5]. In (1.4), of course there is there is an implicit mapping from ordinary into spinor space: if the electron has ordinary space coordinate $(\alpha, \beta)$, then this is also supposed to be the coordinate of its spinor in spinor space. We see that the degree of the polynomial in the skyrmion case is precisely the same as in the Laughlin quasihole case. Thus there is the same, fractional, charge deficit. Notice that the freedom to choose a function in (1) has entirely disappeared.

## 2. Spin and Statistics

The preceding discussion is incomplete: since the state has no definite value of the spin in the up direction, it is embedded in a highly degenerate continuum. Also, there is nothing to vary in the proposed skyrmion wave function, which does not seems reasonable – one expects to be able to consider skyrmions of different size. These problems are related, and both are solved by paying proper attention to the quantization of the collective coordinate corresponding to overall rotation of the anti-skyrmion.

One constructs states of definite angular momentum (and spin) by forming the superposition

$$\Psi_{\text{skyr. } J} = \int_0^{4\pi} d\lambda e^{-iJ\lambda} e^{-iN\lambda/2} \Psi_{\text{skyr. } \lambda} \qquad (2.1)$$

where $\Psi_{\text{skyr. } \lambda}$ is the rotated version of $\Psi_{\text{skyr.}}$, obtained by replacing $\alpha \to e^{i\lambda/2}\alpha$, $\beta \to e^{-i\lambda/2}\beta$ in (1.3). Thus for $J = 0$ one recovers the fully polarized Laughlin quasihole. Thus our skyrmion family constitutes a proper generalization of the traditional *ansatz* for the quasiparticle. For $J = -n$, where $n$ is a natural number, one obtains a state differing from this by flipping $n$ up spins into down spins. The Laughlin quasihole, in a fully polarized $1/m$ state, has number deficit $1/m$ and thus spin-up deficit $1/2m$ relative to the uniform ground state. We are led to conclude that the quasiparticle described by $\Psi_{\text{skyr. } J}$ has spin-up deficit $(1/2m) + n$ relative to the ground state, or effective spin $(-1/2m) + J$.

Notice that the only parameter we get to vary in this construction is $J$. The quantum skyrmion thus exhibits a remarkable rigidity. The question which $J$ is favored for low-lying charged quasiparticles in a given material is a non-universal question, whose answer depends on the detailed form of the Hamiltonian – that is, it involves energetics, not merely topology. Thus one expects to find that the $J$ which minimizes the energy for a quasihole exhibits jumps as one changes the in-plane **B** field or material parameters such as density, impurity concentration, temperature, or well size in the third direction. This effect suggests a method of checking the fractional quantization of the spin. Indeed, using nuclear magnetic resonance one can measure the Knight shift induced by a skyrmion, which is proportional to its spin [6]. If the favored value of the spin jumps by an integer in response to a small change in the control parameters, then by taking the ratio of Knight shifts before and after the change one could infer the ratio, which is of course sensitive to the fractional displacement. In a material that is not perfectly homogeneous, one might find stable skyrmions with different values of $J$ at different positions; and at finite temperature one expects to find each $J$ value represented with appropriate statistical weight.

It remains to discuss one last anomalous quantum number of the skyrmion, that is its anomalous quantum statistics. Given the known result for the traditional Laughlin quasihole, which is the special case $J = 0$, and the results that the fractional part of the spin and charge are independent of $J$, one might anticipate that the statistics also is independent of $J$, and thus for example yields anyons with statistical parameter $\theta = \pi/m$ at filling fraction $\nu = 1/m$, for all possible skyrmions. The "ribbon argument" connecting spin to statistics, whose essence was presented in [7] and adapted to the present context in [2], also leads to this conclusion. It also follows if one works within the effective theory, because the same parameter – that is, the fractional part of the coefficient of the Hopf term – governs all three anomalies. While the anticipated result is true, so far we have not found a really elementary or brief derivation, and here we shall confine ourselves to sketching the framework wherein a formal derivation can be constructed.

A multi-skyrmion *ansatz* can be constructed by generalizing (1.4) in the manner

$$f_{\text{texture}} = \prod \begin{pmatrix} g(\alpha_k(t), \beta_k(t)) \\ h(\alpha_k(t), \beta_k(t)) \end{pmatrix} \qquad (2.2)$$

where $g$ and $h$ are homogeneous complex polynomial functions, say of degree $s$ in their arguments. This will produce a configuration with skyrmion number $s$, although not necessarily with localized charge or spin. (One can also introduce anti-skyrmions, by allowing $\frac{\partial}{\partial \alpha}, \frac{\partial}{\partial \beta}$ as arguments, properly localized.) In any case, one wants to construct an effective action for the texture fields $g, h$. The energy expression will in general be complicated and non-universal, but it will not generate any topological (Hopf) term. Such a term arises from the fact that for the spinor fields to represent spin 1/2 quantum variables, one must regard the coordinate in spinor space as propagating under the influence of a Dirac monopole vector potential. One can demonstrate mathematically that including these factors in evaluating the amplitude for the series of configurations parameterized by $(g(t), h(t))$ which begins and ends trivially generates a phase proportional to the Hopf invariant of the corresponding map $S^3 \to S^2$, with the anticipated coefficient. Thus, we have the following effective Lagrangian for the local spin field, $n_a$:

$$\mathcal{L} = \mathcal{L}_0 - \frac{4\pi}{m}\left(j^\alpha a_\alpha - \frac{1}{2}\epsilon^{\alpha\beta\gamma}a_\alpha \partial_\beta a_\gamma\right) \tag{2.3}$$

where $j^\alpha = \frac{1}{8\pi}\epsilon^{\alpha\beta\gamma}\epsilon^{abc}n_a\partial_\beta n_b \partial_\gamma n_c$ is the skyrmion current, $a_\alpha$ is a gauge field which can be integrated out to give a non-local term in terms of $n_a$ alone, and $\mathcal{L}_0$ is the non-linear $\sigma$-model Lagrangian discussed in [4] with gradient energy, Zeeman energy, and Coulomb repulsion terms. We can redefine $\tilde{a} = \frac{4\pi}{m}a$, so that the Chern-Simons term is conventionally normalized and the quantized parameter appears explicitly as a coefficient in the Lagrangian. Its quantization is connected with the invariance of the action under large gauge transformations [8]. Given the mathematical result (2.3), the ribbon argument of [2] applies, and the anyon character of the skyrmion follows.

To put this formal development into context, two remarks are appropriate. First, the structure of the skyrmions themselves cannot be cleanly derived within the effective field theory, except for $J \gg 1$ – although their long-wavelength interactions, including especially their quantum numbers, can be summarized in it. Physically, this is because the characteristic length scale of the skyrmions is comparable to the magnetic length, for small $J$, and there is no small expansion parameter for a gradient expansion; mathematically, it is because the restriction to holomorphic functions (characteristic of the lowest Landau level) is not captured, at least within a conventional $\sigma$-model.* On the other hand, just because of this rigidity isolated skyrmions have an energy gap (apart from the rotational zero mode) and therefore use of microscopic trial wavefunctions is less questionable than is usually the case for many-body problems.

## 3. Skyrmion Condensation and the Hierarchy Construction

The exotic spin of the skyrmions allows us to understand spin-singlet states and, more generally, non-polarized states as hierarchical states resulting from the condensation of skyrmionic quasiparticles on a polarized parent state. To see why this is non-trivial, recall that, in the hierarchy construction, the state at $\nu = 2/5$ forms when charge $-e/3$ and statistics $-\pi/3$ quasiparticles of the polarized $\nu = 1/3$ state condense in a Laughlin state. An alternative viewpoint, the flux-trading procedure, which relates states at $\nu$ and $\frac{\nu}{2p\nu \pm 1}$ implies that $\nu = 2/5$ shares many qualitative features with $\nu = 2$. Since a spin-singlet or spin-polarized state could form at $\nu = 2$, depending on the ratio of Zeeman and cyclotron energies, we expect both possibilities at $\nu = 2/5$. There is numerical and experimental evidence that this is correct. But how can a spin-singlet state ever descend from a polarized $\nu = 1/3$ state when this would mean that the additional $\nu = \frac{2}{5} - \frac{1}{3} = \frac{1}{15}$ must cancel the spin of the $\nu = 1/3$ parent?

The observation that skyrmions are the lowest energy quasiparticles in the low Zeeman energy limit at $\nu = 1/3$ is the key. The anomalous spin of the skyrmions is crucial for this picture. Since skyrmions have the same charge and statistics as the Laughlin quasiparticles, the allowed fractions are the same as for the spin-polarized hierarchy. In other words, the state at $\nu = 2/5$ is a state with one-third of an electron and one-fifth of a skyrmion per flux tube. The electrons are spin-polarized and carry spin $S_z = \frac{1}{2}$. The skyrmions have, by the arguments of the previous sections, spin $S_z = J - \frac{1}{6}$ oppositely directed to the electron spin. Then, the total $S_z$ per flux tube is $S_z = \frac{1}{2} \times \frac{1}{3} - \left(J - \frac{1}{6}\right) \times \frac{1}{5}$. If $J = 1$, $S_z = 0$. More generally, a daughter state in which skyrmions of a Laughlin state condense will have charge and spin filling fraction:

$$\nu = \frac{1}{m} + \frac{\alpha}{m} \frac{1/m}{2p - \alpha/m} \qquad (3.1)$$

$$S_z = \frac{1}{2} \times \frac{1}{m} - \left(J - \frac{\alpha}{2m}\right) \times \frac{1/m}{2p - \alpha/m} \qquad (3.2)$$

where $\alpha = \pm 1$ according to whether skyrmions or anti-skyrmions condense. Observe that the fractional part of the spin is either aligned or anti-aligned with the parent, depending on whether it is particle- or hole-like ($\alpha = \pm 1$), but the integer part is always anti-aligned because it involve flipping spins of the parent condensate. This state will have $S_z = 0$ if $J = p$. It is natural that the most favorable skyrmion size, $J$, be determined by the skyrmion inverse density, $p$, in the low Zeeman energy, high-density limit, where inter-skyrmion interactions are the limiting factor. Indeed, the existence of spin-singlet ground states at the $p = 1$ fractions (at least numerically) implies that $J = 1$ has the lowest energy when $p = 1$. Still, it would be remarkable if the relation between the energetically favored $n$ and $p$ holds in general. The skyrmion condensation picture of the states at $\nu = \frac{2p}{2pm\pm 1}$ motivates the following trial wavefunction for these states:

$$\Psi_{\frac{2p}{2pm\pm 1}} = \int \prod_l D\gamma_l D\delta_l \prod_{i>j}(\gamma_i \delta_j - \gamma_j \delta_i)^{2p} \prod_r f_{\text{s,as}}^J(\gamma_r, \delta_r) \Psi_m \qquad (3.3)$$

where $f_{\text{s,as}}(\gamma_r, \delta_r)$ is the factor for a skyrmion or anti-skyrmion of spin $J \pm \frac{1}{2m}$ centered at $(\gamma_r, \delta_r)$. If the variational energy of this wavefunction is minimized for $J = p$ (low Zeeman energy limit) then the ground state has $S_z = 0$ but if it is minimized for $J = 0$ (high Zeeman energy limit), the ground state is polarized. The transition between these ground states is then just a transition between the favored $J$'s which is necessarily first-order. Other spin-polarized states would have partially polarized descendents (but never $S_z = 0$) via skyrmion condensation. Finally, we would expect states in which skyrmions form a Fermi sea at certain even-denominator fractions, such as $\nu = 1/2$ ($\alpha = -1$, $m = 1$, $2p = 1$ in (3.2)) and $\nu = 3/4$ ($\alpha = -1$, $m = 1$, $2p = 3$). In all of the above considerations, it has been absolutely necessary for the skyrmions to have large spin with the fractional part given by the spin-statistics theorem.

---

* However, it is an intriguing coincidence that the classical skyrmions solutions of the non-linear $\sigma$-model (2.3) are parameterized by an analytic function [9], just as the lowest Landau level wavefunctions are.